\documentclass[10pt]{article}
\usepackage{graphicx}

\def\Title#1{\begin{center} {\Large #1 } \end{center}}
\def\Author#1{\begin{center}{ \sc #1} \end{center}}
\def\Address#1{\begin{center}{ \it #1} \end{center}}

\newcommand\pubblock{\rightline{\begin{tabular}{l} Proceedings of the Fifth Annual LHCP\\ \pubnumber\\
         \pubdate  \end{tabular}}}

\newenvironment{Abstract}{\begin{quotation} \begin{center} 
             \large ABSTRACT \end{center}\bigskip 
      \begin{center}\begin{large}}{\end{large}\end{center} \end{quotation}}

\newenvironment{Presented}{\begin{quotation} \begin{center} 
             PRESENTED AT\end{center}\bigskip 
      \begin{center}\begin{large}}{\end{large}\end{center} \end{quotation}}

\def\beq{\begin{equation}}
\def\eeq#1{\label{#1}\end{equation}}
\def\eeqn{\end{equation}}

\def\beqa{\begin{eqnarray}}
\def\eeqa#1{\label{#1}\end{eqnarray}}
\def\eeqan{\end{eqnarray}}

\let\bar=\overbar

\def\Dslash{\not{\hbox{\kern-4pt $D$}}}
\def\dslash{\not{\hbox{\kern-2pt $\del$}}}

\def\msb{{\bar{\ssstyle M \kern -1pt S}}}

\usepackage{color}

\newcommand\pubnumber{ CMS-CR-2017-196 }

\newcommand\pubdate{\today}

\def\affiliation{
On behalf of the CMS and ATLAS Collaborations, \\
Physics Institute IIIA \\
RWTH Aachen University, Aachen, Germany}

\begin{document}

\large
\begin{titlepage}
\pubblock

\vfill
\Title{ High mass searches in CMS and ATLAS }
\vfill

\Author{ Swagata Mukherjee }
\Address{\affiliation}
\vfill
\begin{Abstract}

The latest results of high mass searches for new physics in a variety of final states from the CMS and ATLAS collaborations
are presented. These searches are based on $\sqrt{s}=13$ TeV proton-proton collisions data at the LHC collected in the year 2016 and 2015. 
No excess above expectation from
Standard Model processes are observed and exclusion limits are set at the 95\% confidence level on various benchmark models.

\end{Abstract}
\vfill

\begin{Presented}
The Fifth Annual Conference\\
 on Large Hadron Collider Physics \\
Shanghai Jiao Tong University, Shanghai, China\\ 
May 15-20, 2017
\end{Presented}
\vfill
\end{titlepage}
\def\thefootnote{\fnsymbol{footnote}}
\setcounter{footnote}{0}
%

\normalsize 


\section{Introduction}
The Standard Model (SM) of particle physics is a remarkably successful theory, with experimental
results being consistently in agreement with its predictions. It is however well-known
that the SM does not describe the nature completely and it is usually seen as a low energy
approximation of a more general theory. Therefore there exist many theories extending beyond
the SM (BSM) which predict new particles at the TeV mass scale. 
After the discovery of the Higgs boson in 2012, the primary goal of the CMS~\cite{Chatrchyan:2008aa} and ATLAS~\cite{Aad:2008zzm} experiments is to discover new BSM physics.
Both these experiments have a rich program of searches for new exotic particles. 

The LHC had a spectacular performance in 2016, exceeding expectations, reaching an instantaneous
luminosity of $1.5 \times 10^{34}$ $\mathrm{cm}^{-2}\mathrm{sec}^{-1}$. The high instantaneous luminosity comes
at the price of pileup, the number of overlapping proton-proton interactions in one bunch crossing. 
In 2016, an average of 24 pileup interactions was observed. However the experiments have
developed tools to mitigate the effect of pileup at the trigger and object reconstruction level.
The ATLAS and CMS experiments collected data with more than 92\% efficiency and approximately 36 fb$^{-1}$ of good quality data was recorded in 2016. 
New results of high mass searches at $\sqrt{s}=13$ TeV with highest available luminosities from the 2016 or 2015 datasets are presented in this report.
\section{Signature-based resonance searches}
Signature-based resonance searches in variety of final states and many different physics interpretations are discussed in this section.
\subsection{Dijet}
Many models of physics beyond the SM require new particles that couple to quarks ($q$) and gluons ($g$) 
\begin{figure}[htb]
\centering
\includegraphics[height=2.3in]{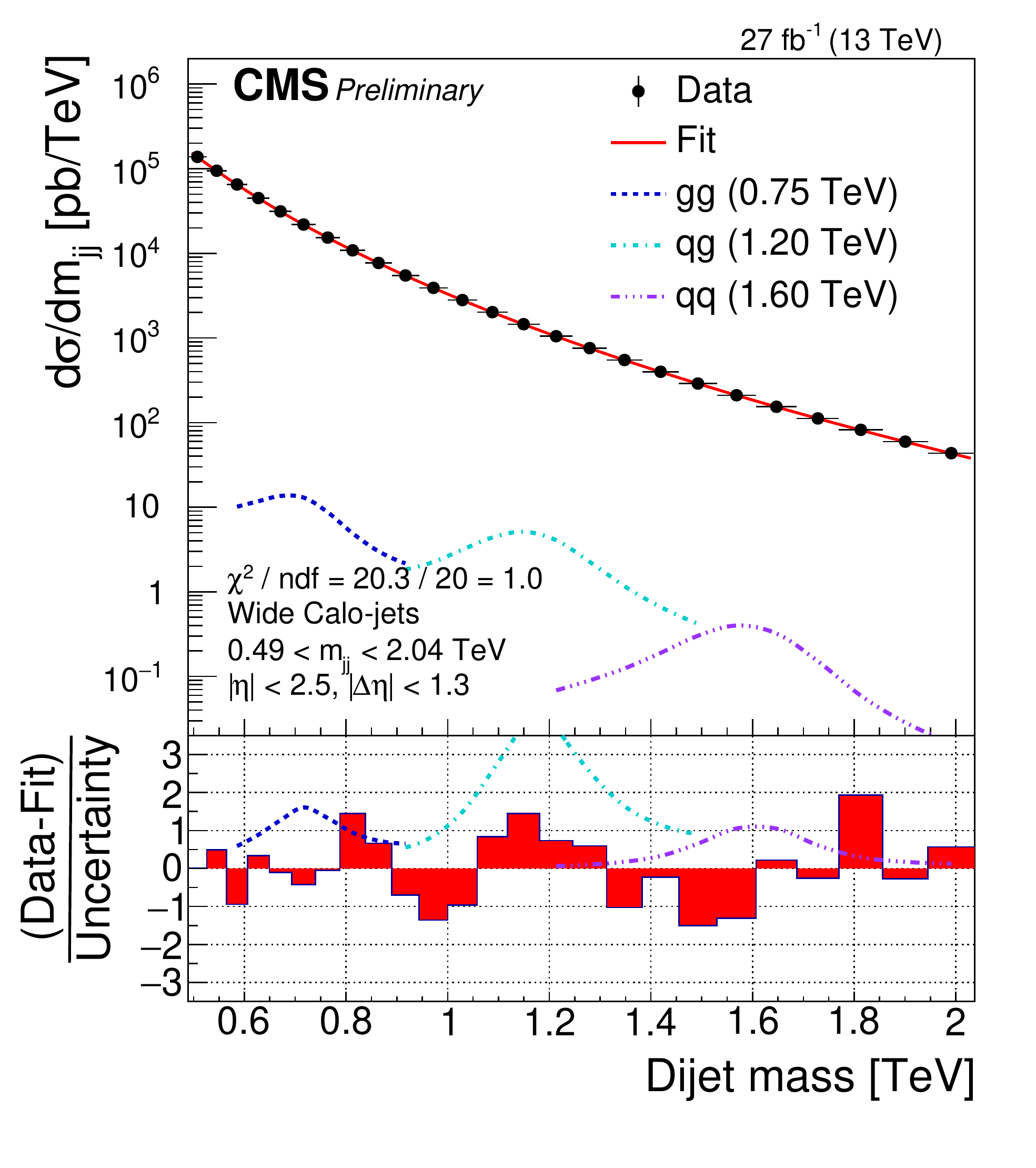}
\hskip 10 pt
\includegraphics[height=2.3in]{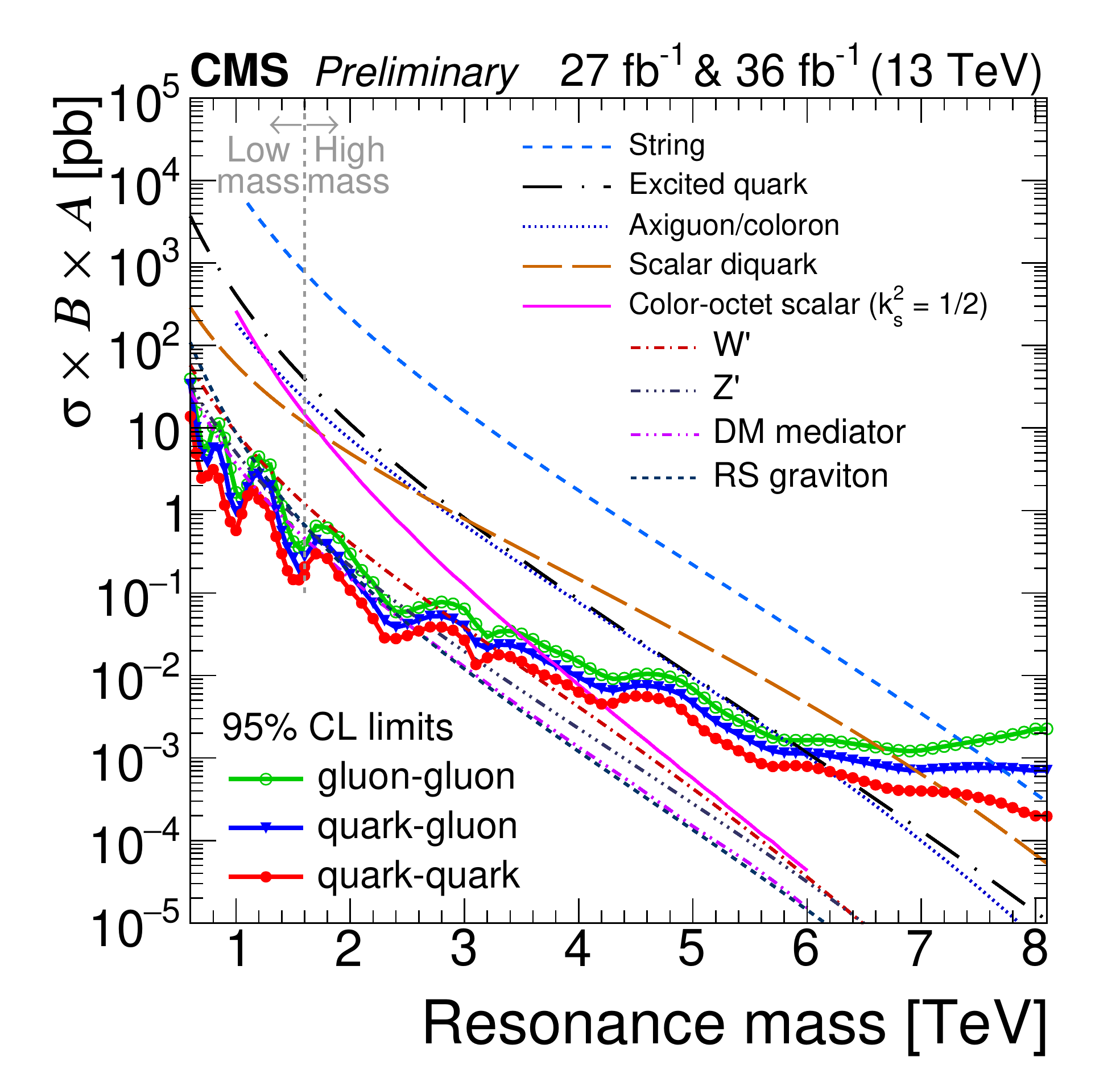}
\caption{Left: Dijet mass spectrum~\cite{CMS:2017xrr} compared to a fitted parameterization of the background (solid red curve) for the low-mass search using scouting technique.  
The lower panel shows the difference between the data and the fitted parametrization, divided by the statistical uncertainty of the data. 
Examples of predicted signals from narrow gluon-gluon, quark-gluon, and quark-quark resonances are shown with cross sections equal to the observed upper limits at 95\% CL. Right: The observed 95\% CL 
upper limits~\cite{CMS:2017xrr} on the product of the cross section, branching fraction and acceptance for quark-quark, quark-gluon, and gluon-gluon type dijet resonances. 
Limits are compared to predicted cross sections for string resonances, excited quarks, axigluons, colorons, scalar diquarks, color-octet scalars, new gauge bosons $W'$ and $Z'$ with SM-like couplings, dark matter mediators for $m_{\mathrm{DM}}=1$ GeV and RS gravitons. }
\label{fig:dijet}
\end{figure}
and can be observed as resonances in the dijet mass spectrum. 
For CMS, a high-mass search~\cite{CMS:2017xrr} for resonances with mass above 1.6 TeV is performed using dijets reconstructed 
with the particle flow algorithm using data corresponding to an integrated luminosity of 36 fb$^{-1}$ at 13 TeV. A low-mass search is also 
performed for resonances with mass between 0.6 and 1.6 TeV using dijet events corresponding to an integrated luminosity of 27 fb$^{-1}$ at 13 TeV, where the events are reconstructed, selected, and recorded in a compact form by the high-level trigger (HLT) in a technique called data scouting. Figure~\ref{fig:dijet}-left shows the dijet mass spectrum for the low-mass dijet search.
In the analyzed data samples, there is no evidence for resonant particle production. Generic upper limits are presented on the product of the cross section, the branching fraction, and the acceptance for narrow quark-quark, quark-gluon, and gluon-gluon resonances, as shown in Figure~\ref{fig:dijet}-right.
In the context of specific models, the limits exclude string resonances with masses below 7.7 TeV, scalar diquarks below 7.2 TeV, axigluons and colorons below 6.1 TeV, 
excited quarks below 6.0 TeV, color-octet scalars below 3.4 TeV, $W'$ bosons below 3.3 TeV, $Z'$ bosons below 2.7 TeV, 
RS gravitons below 1.7 TeV and between 2.1 and 2.5 TeV, and dark matter mediators below 2.6 TeV. 
Similar searches are also performed by ATLAS collaboration and can be found in Refs.~\cite{Aaboud:2017yvp,ATLAS:2016gvq,ATLAS:2016xiv}. 
%
%
%
\subsection{Dijet $+$ ISR}
\begin{figure}[htb]
\centering
\includegraphics[height=2.2in]{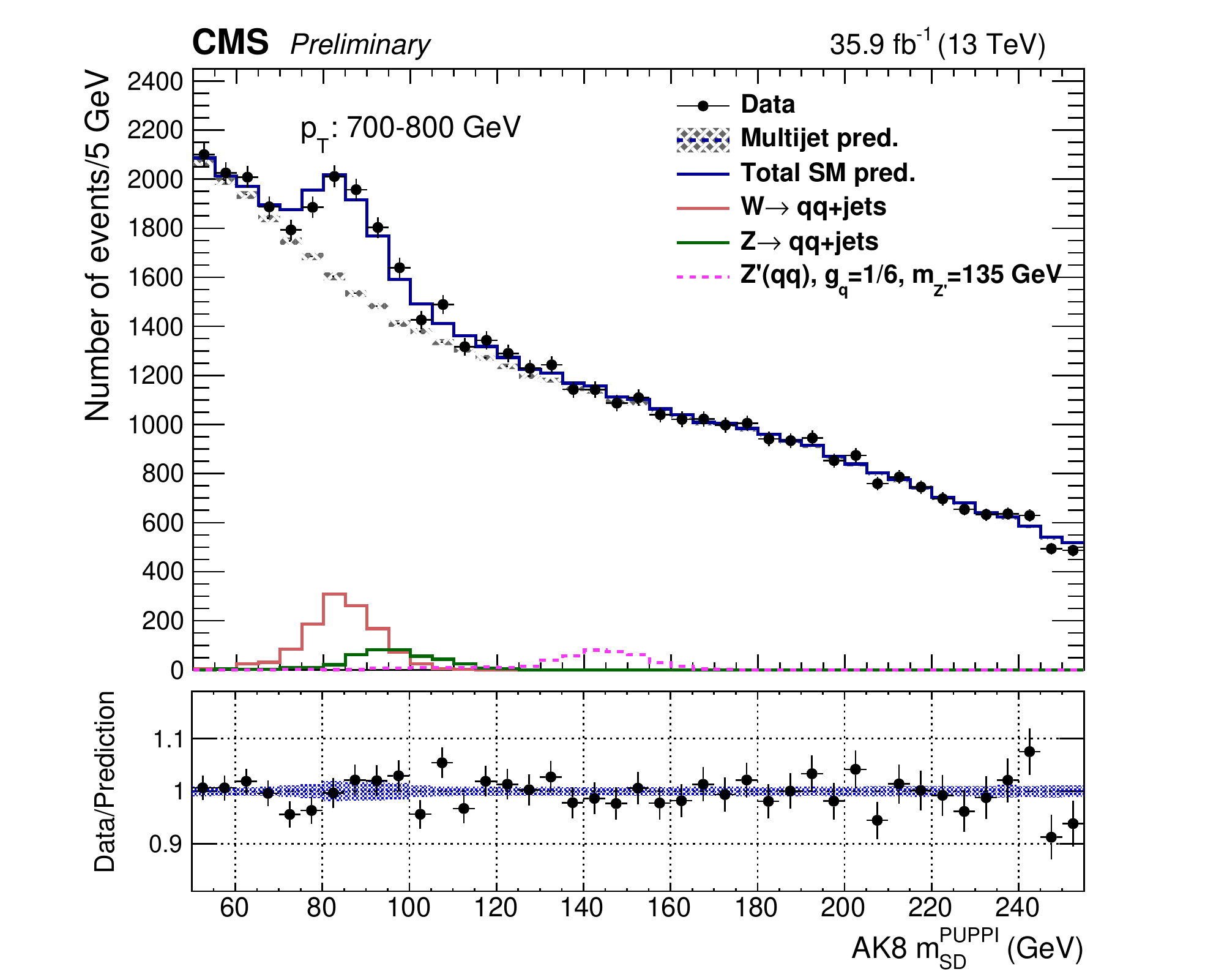}
\hskip 10 pt
\includegraphics[height=2.2in]{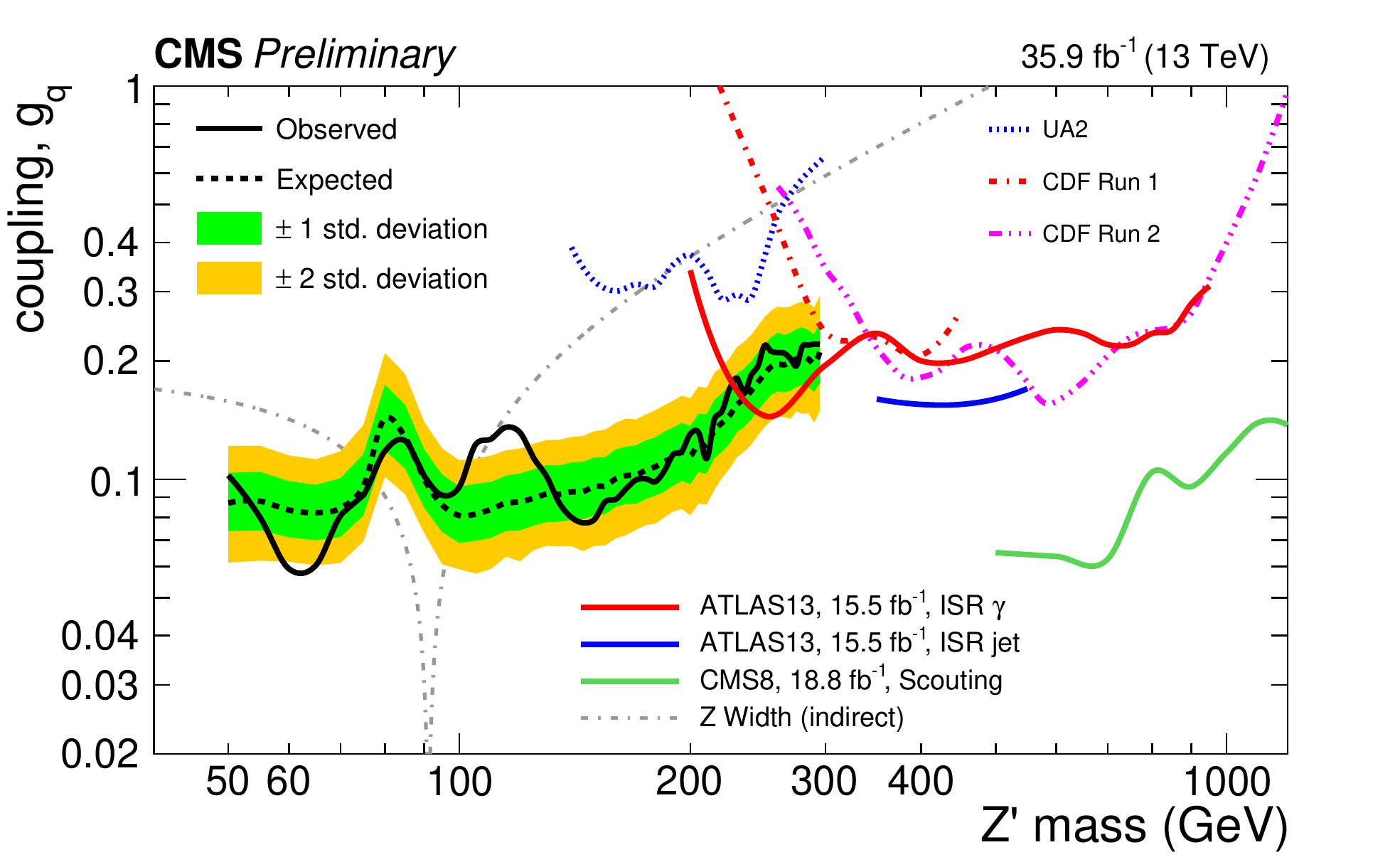}
\caption{Left: Soft drop jet mass distribution~\cite{CMS:2017dhi} for the $p_{\mathrm{T}}$ category of 700-800 GeV. Data are shown as the black points. 
The QCD background prediction, including uncertainties, is shown in the gray boxes. Contributions from the $W$, $Z$, and a hypothetical $Z'$ signal at a mass of 135 GeV are indicated as well. 
Right: Limits on the coupling $g_q$ as a function of the leptophobic $Z'$ mass~\cite{CMS:2017dhi}. Limits from other relevant searches are also shown. 
An indirect constraint on a potential $Z'$ signal from the SM $Z$ boson width is also shown.}
\label{fig:dijetISR}
\end{figure}

At resonance masses below 0.6 TeV the sensitivity of the dijet search is limited both by the large SM multijet production rate and by trigger and storage requirements. 
These difficulties can be avoided by an alternative approach in which a high 
transverse momentum jet from initial-state radiation (ISR) is required to be produced in association with the light diquark resonance. 
The ISR jet constraint provides enough energy in the event to satisfy the trigger requirements. In CMS search~\cite{CMS:2017dhi}, 
background jet combinatorics are reduced by assuming that the 
dijet resonance is sufficiently boosted, so that the hadronization products merge and are reconstructed within a single massive jet. 
Soft drop jet mass distribution for one of the five different $p_{\mathrm{T}}$ categories used in this analysis is shown in figure~\ref{fig:dijetISR}-left.
No evidence for such resonance is observed within the targeted mass range of 50-300 GeV.
Upper limits at a 95\% confidence level are set on the production cross-section of leptophobic vector resonances. 
Results are also presented in a mass-coupling phase space, as shown in Figure~\ref{fig:dijetISR}-right, and are the most sensitive to date, extending previous limits below 100 GeV~\cite{Sirunyan:2017dnz}.

A similar search, requiring the new dijet resonance to be produced in association with a high-$p_{\mathrm{T}}$ photon or jet, is performed by 
ATLAS experiment~\cite{ATLAS:2016bvn}. This search is performed in resolved dijet regime and probes 200-1500 GeV for the $X+\gamma$ search and 
300-600 GeV for the $X + j$ search, where $X$ is the new resonance decaying to dijet.   
%
\subsection{Pair-produced dijets}
A search for the pair production of resonances, each decaying into two jets, has been performed by ATLAS collaboration 
using 15.4 fb$^{-1}$ of proton-proton collision data recorded at $\sqrt{s}=13$ TeV~\cite{ATLAS:2016sfd}.
\begin{figure}[htb]
\centering
\includegraphics[height=2.3in]{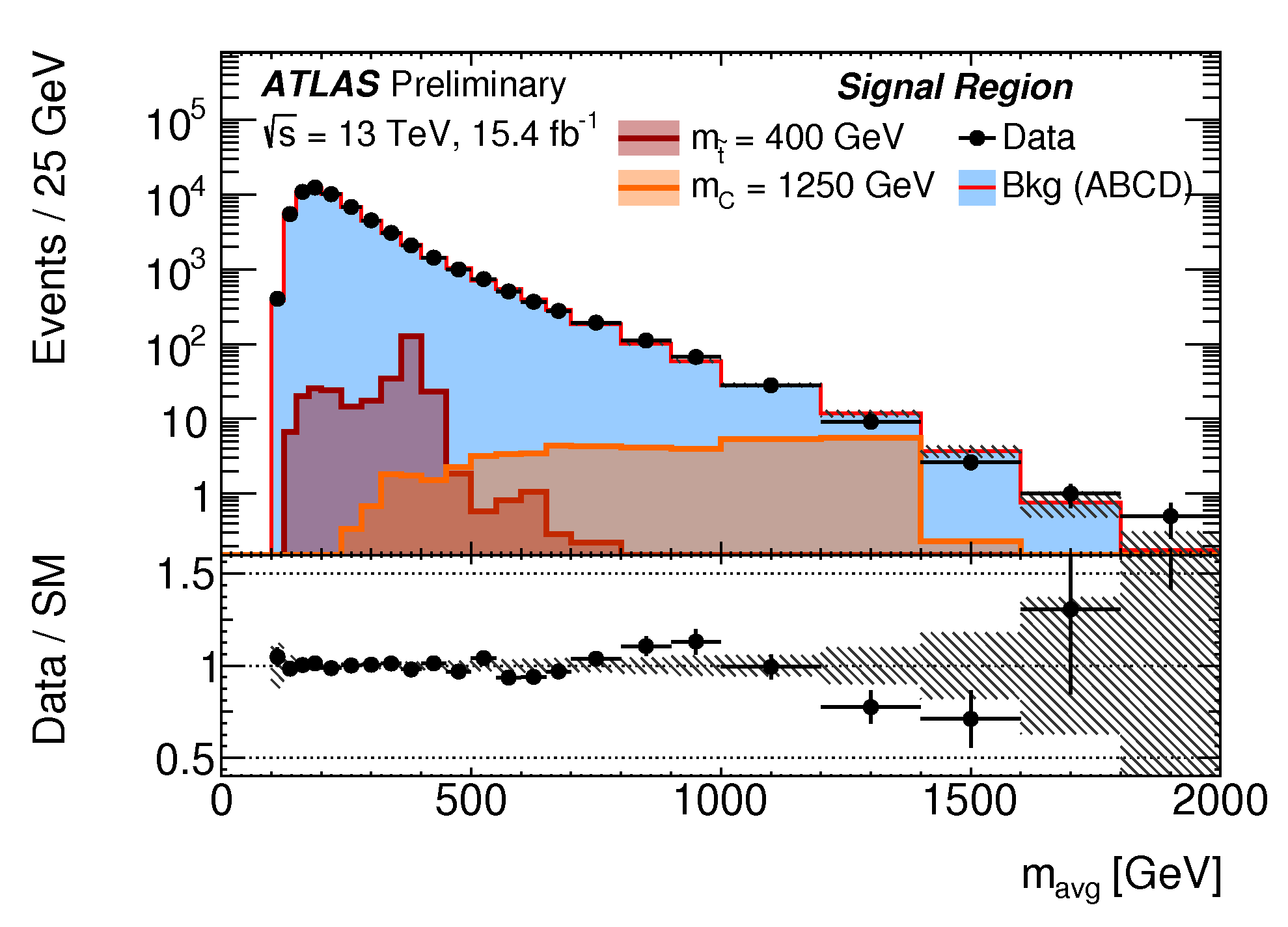}
\includegraphics[height=2.5in]{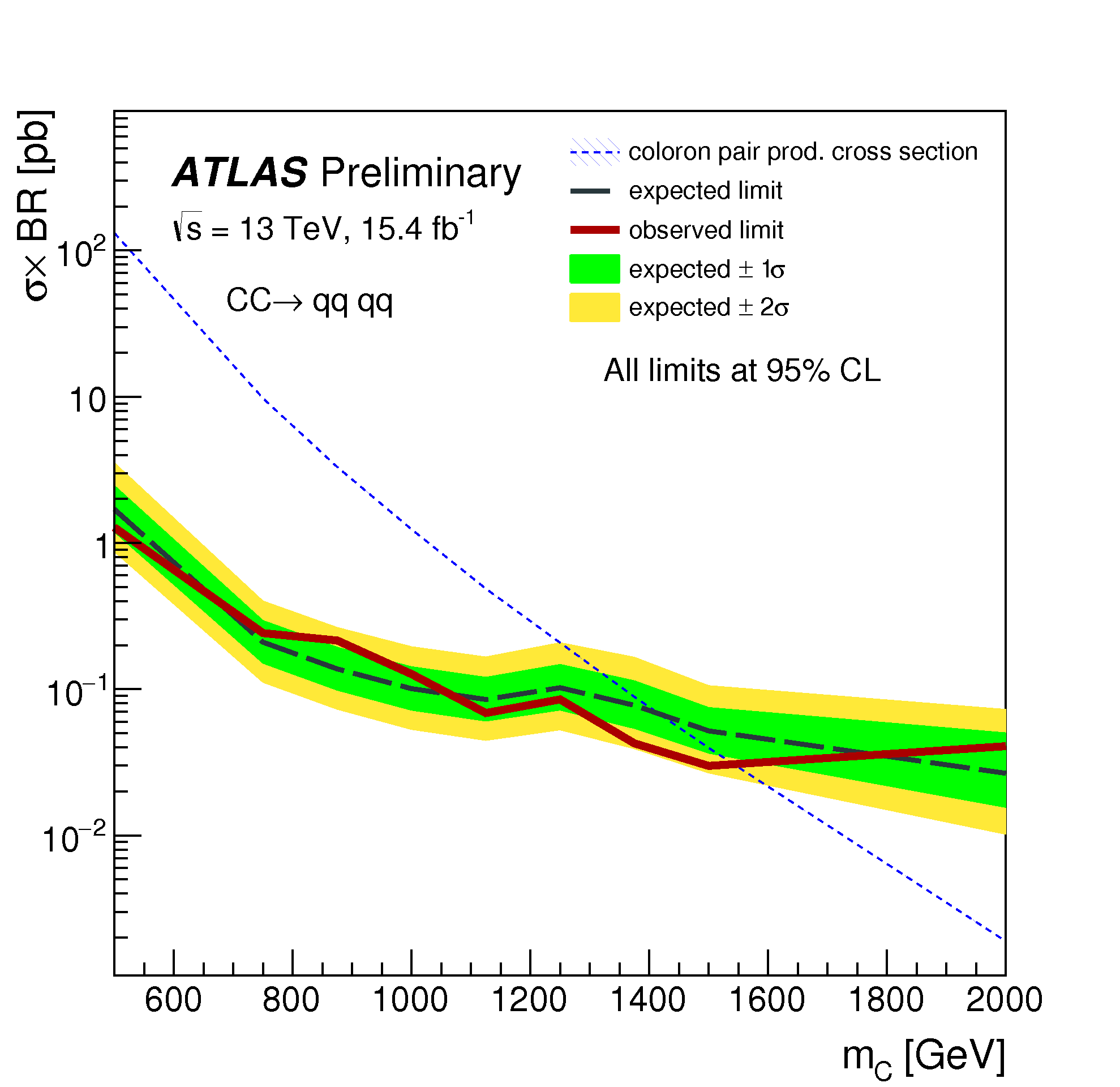}
\caption{Left: The observed $m_{\mathrm{avg}}$ spectrum~\cite{ATLAS:2016sfd} in the signal region (black points) compared to 
the total background prediction (blue histogram) estimated with a data-driven method. In the bottom pad, all 
systematic uncertainties on the background estimate are included in the grey hatched band. The expected distributions for a few representative signal points are overlaid.
Right: Expected and observed 95\% CL upper limits~\cite{ATLAS:2016sfd} on the $\sigma \times$BR for the signal, compared to the theoretical cross-section for coloron production with decays to a pair of jets.}
\label{fig:4jet}
\end{figure}
The final analysis discriminant is the average mass of the two reconstructed resonances: 
$$
m_{\mathrm{avg}}=\frac{1}{2}(m_1 + m_2)
$$
The $m_{\mathrm{avg}}$ distribution, after applying all selection cuts, is shown in Figure~\ref{fig:4jet}-left.
No significant excess is observed above the background prediction. The results are interpreted in a model with top squark pair production with R-parity violating decays into two quarks. Top squark masses between 250 and 405 GeV and between 445 and 510 GeV are excluded at 95\% confidence level. 
For the pair production of color octets with decays into two jets, masses from 250 to 1500 GeV are excluded at 95\% confidence level, as shown in Figure~\ref{fig:4jet}-right.
CMS experiment performed a similar search in boosted regime utilizing jet substructure techniques, using 2.7 fb$^{-1}$ of 13 TeV data~\cite{CMS:2016pkl}. 
This search probes the mass range of 80-300 GeV.
%
%
\subsection{Dilepton}
\begin{figure}[htb]
\centering
\includegraphics[height=2.3in]{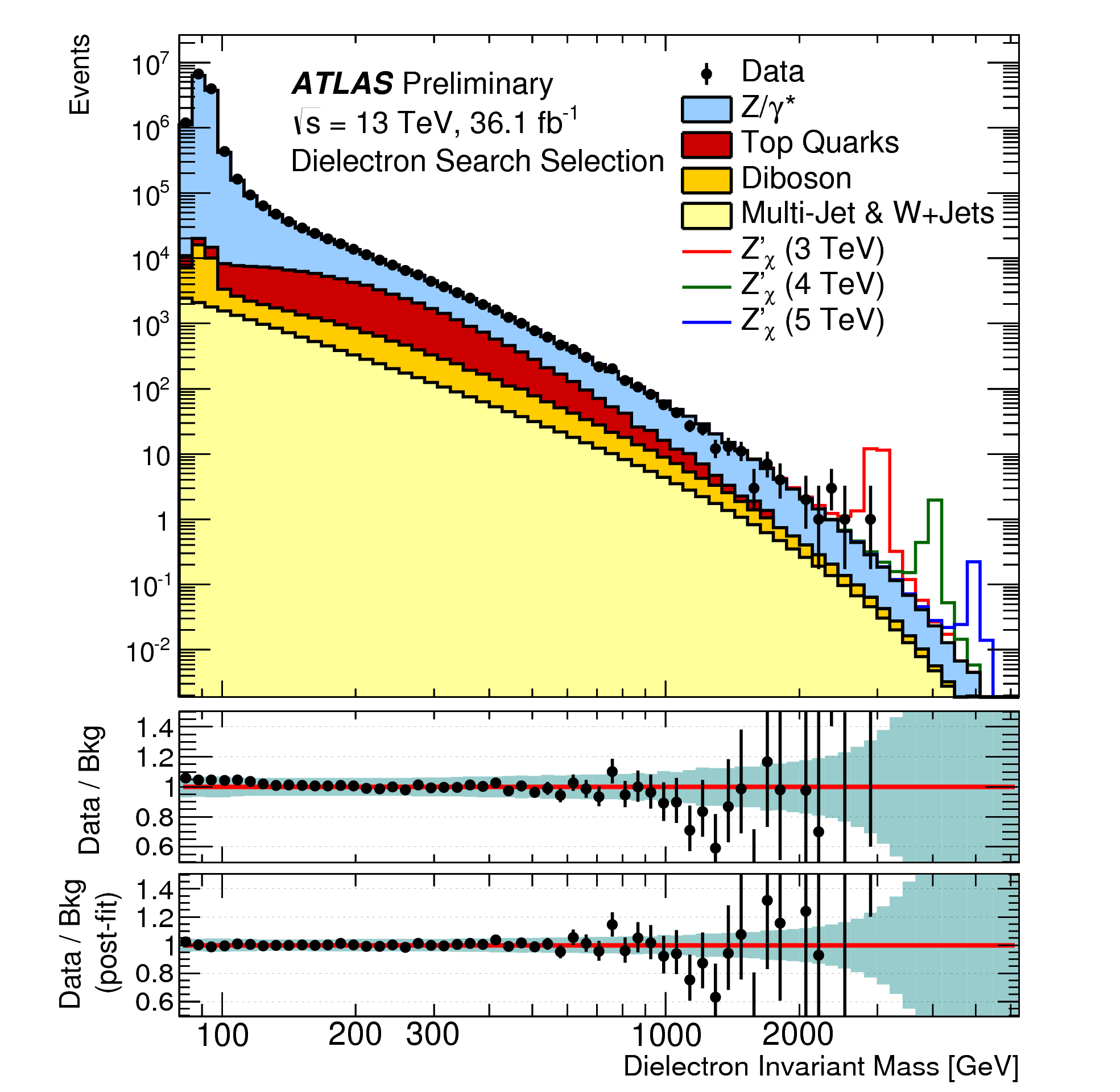}
\hskip 10 pt
\includegraphics[height=2.3in]{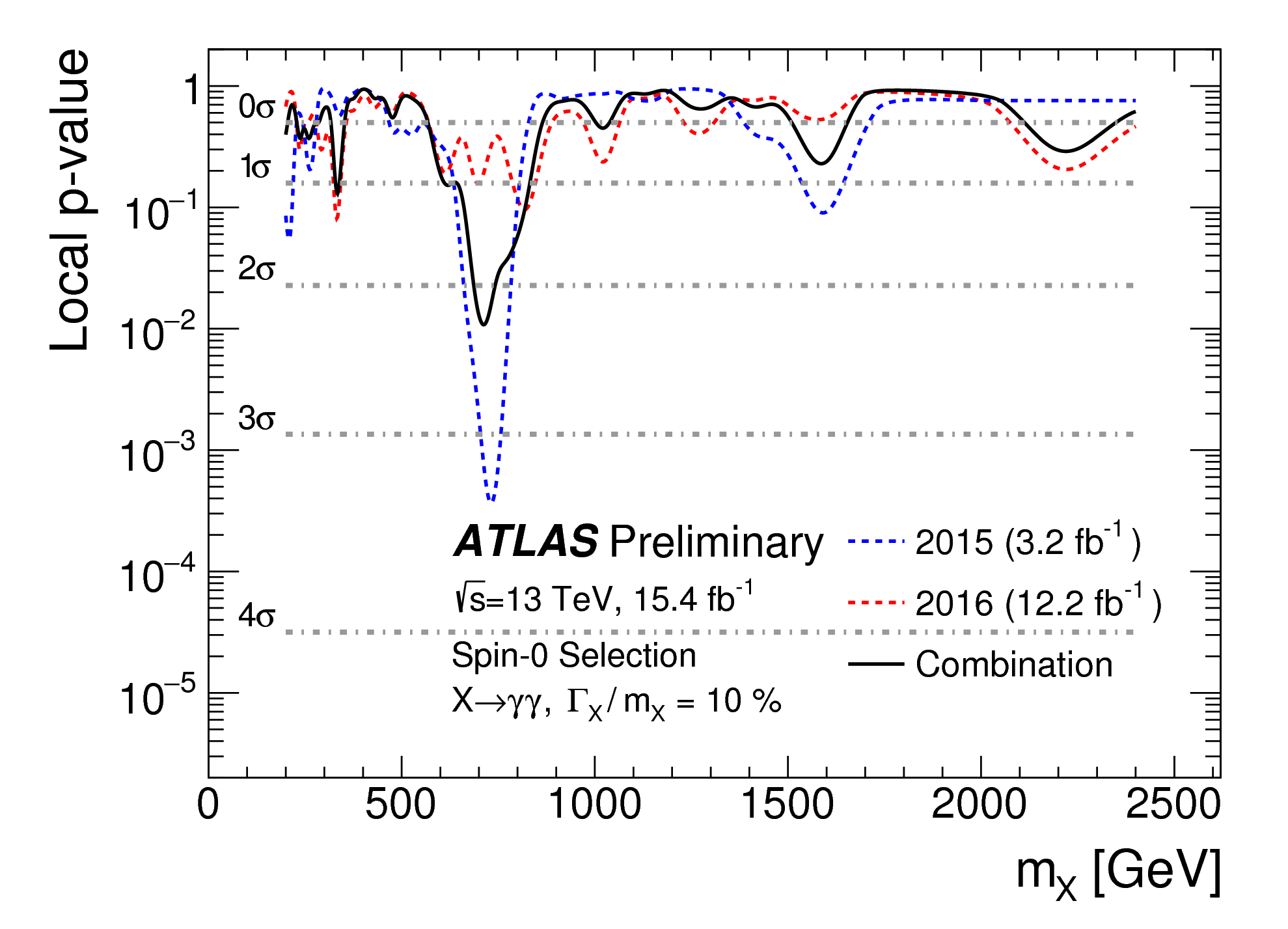}
\caption{Left: Distribution of dielectron reconstructed invariant mass~\cite{ATLAS:2017wce} after selection, for data and the SM background estimates as well as their ratio before and after marginalisation.
Selected $Z'_{\chi}$ signals with a pole mass of 3, 4 and 5 TeV are overlaid. The shaded band in the lower panels illustrates the total systematic uncertainty.
Right: Compatibility with the background-only hypothesis as a function of the assumed signal mass for 10\% relative width $\Gamma_X/m_X$ of a spin-0 resonance decaying to diphoton~\cite{ATLAS:2016eeo}.}
\label{fig:ll_gg}
\end{figure}

The study of dilepton final-states provides excellent sensitivity to a large variety of phenomena. 
This experimental signature benefits from a fully reconstructed final state, high signal selection efficiencies and relatively small, 
well-understood backgrounds, representing a powerful test for a wide range of theories beyond the Standard Model.
Models with extended gauge groups often feature additional $U(1)$ symmetries with corresponding heavy, spin-1, neutral boson, 
generally referred to as $Z'$, the decay of which would manifest itself as a narrow resonance
in the dilepton mass spectrum. The ATLAS collaboration performed a search for new resonant and 
non-resonant high-mass phenomena in dielectron and dimuon final states using 36.1 fb$^{-1}$ 
of proton-proton collision data~\cite{ATLAS:2017wce}. 
The distribution of the dielectron invariant mass
is shown in Figure~\ref{fig:ll_gg}-left.
The highest invariant mass event is found at 2.9 TeV.
No significant deviation from the SM prediction is observed. Upper limits at 95\% credibility level 
are set on the cross-section times branching ratio for resonances decaying to dileptons, which are converted into lower limits on the resonance mass, up to 4.1 TeV for the 
$E_6$-motivated $Z'_{\chi}$, and 4.5 TeV for the $Z'_{\mathrm{SSM}}$ arising in Sequential Standard Model.
Lower limits on the $llqq$ contact interaction scale are set between 23.5 TeV and 40.1 TeV, depending on the models.
A similar search is performed by CMS collaboration and can be found in Ref.~\cite{CMS:2016abv}.
%
%
%
\subsection{Diphoton}
Given fairly strong limits set on the masses of new resonances using fermionic decay channels (e.g. dijet or dilepton searches), 
it is particularly interesting to explore bosonic decay channels, which can dominate if the coupling to fermions is suppressed.
Many extensions to the SM predict new spin-0 or spin-2 resonances that decay to a diphoton final state, which 
provides a clean experimental signature with excellent invariant mass resolution and moderate backgrounds.
Using 3.2-3.3 fb$^{-1}$ of 13 TeV data recorded in 2015, the ATLAS and CMS collaborations reported an excess in the diphoton invariant mass spectra with respect to the SM continuum background near the mass value of 750 GeV. This excess was not confirmed in 2016 data.

A search for new spin-0 resonances decaying into two photons is performed by the ATLAS experiment based on $pp$ collision 
data corresponding to an integrated luminosity of 15.4 fb$^{-1}$ at 13 TeV~\cite{ATLAS:2016eeo}. 
No significant excess is observed in the 2016 data. 
Figure~\ref{fig:ll_gg}-right illustrates the local p-value as a function of diphoton mass for 10\% resonance width, comparing the results observed with the 2015 dataset only, the 2016 dataset only and the combined dataset.
In the absence of any significant excess, limits on the production cross section times branching ratio to two photons of spin-0 resonances are reported in Ref.~\cite{ATLAS:2016eeo}.

A similar search is performed by CMS experiment using 12.9 fb$^{-1}$ of 13 TeV data recorded in 2016 and can be found in Ref.~\cite{Khachatryan:2016yec}. 
No significant excess was found in the 2016 analysis.
The results of the search are combined statistically with those previously obtained in 2012 and 2015 at $\sqrt{s}=8$ and 13 TeV, respectively, corresponding to 
integrated luminosities of 19.7 and 3.3 fb$^{-1}$. The combined analysis excludes RS gravitons with masses below 3.85 and 4.45 TeV for the dimensionless coupling parameter
$\tilde{\kappa}=0.1$ and 0.2, respectively.  

\subsection{Z$\gamma$}
While a resonance decaying to diphotons cannot be a vector or an axial vector, due to the Landau-Yang theorem, having one of the two 
bosons massive diminishes this constrain. 
\begin{figure}[htb]
\centering
\includegraphics[height=2.3in]{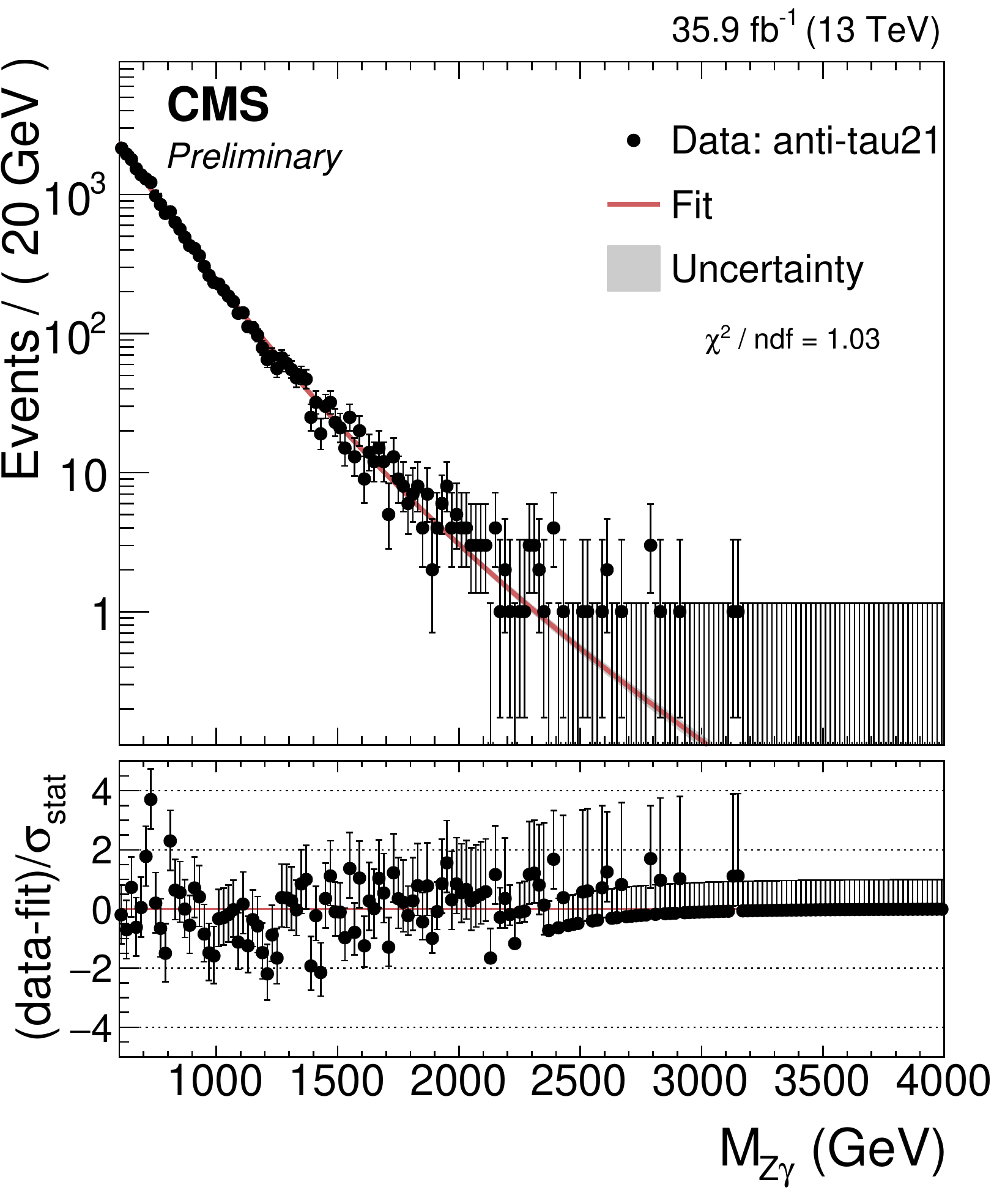}
\hskip 10 pt
\includegraphics[height=2.4in]{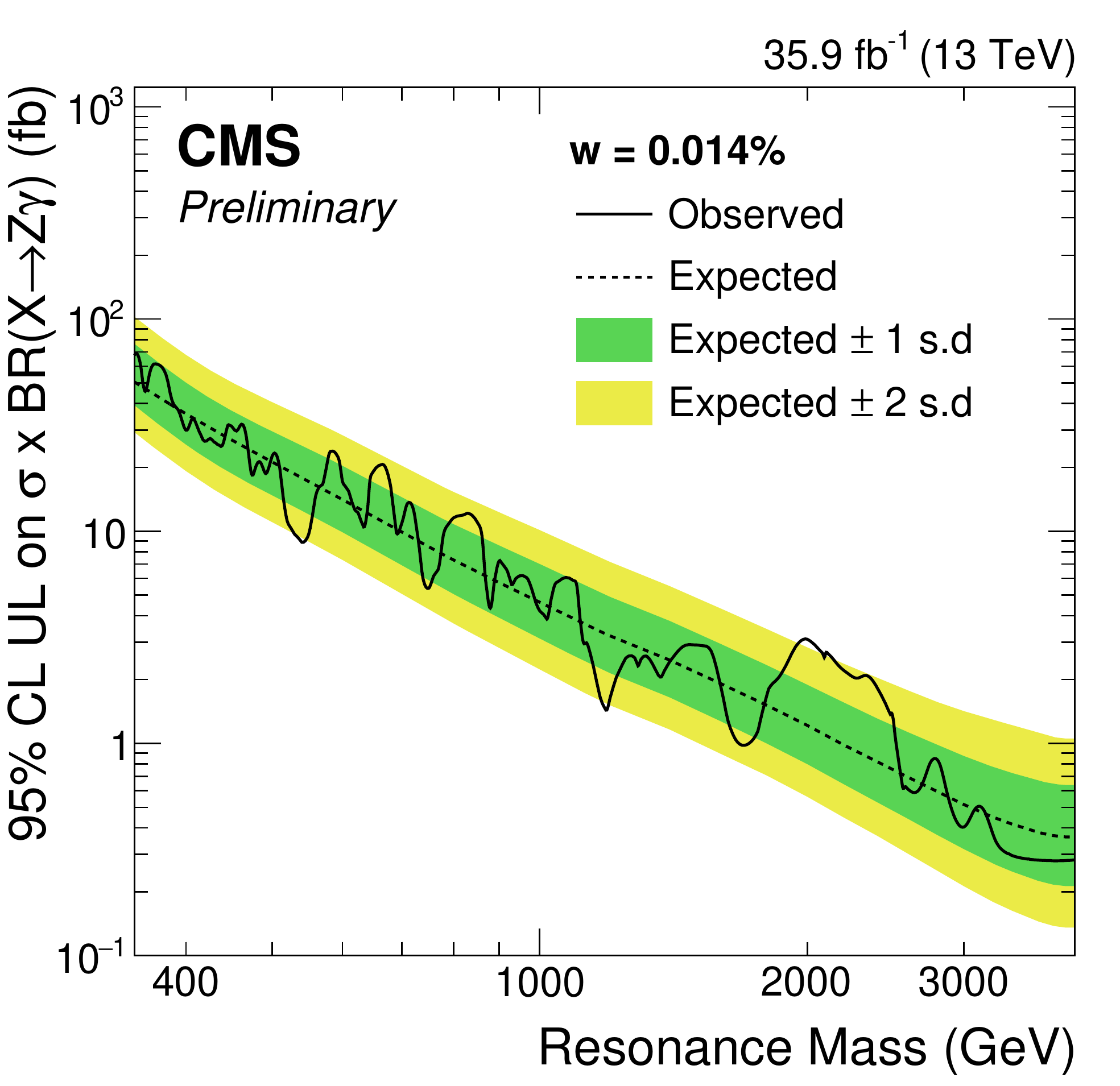}
\caption{Left: Distribution of $M_{Z\gamma}$ and corresponding fit in the signal region in the anti-tau21 category~\cite{CMS:2017fge} in the hadronically decaying $Z$ plus $\gamma$ analysis. 
Right: Observed and expected limits~\cite{CMS:2017fge} on the product of the cross section and branching fraction $B(X \rightarrow Z\gamma)$ for the 
production of a narrow spin-0 resonance, obtained from the combination of the 13 TeV analyses in hadronic and leptonic decay channels of the $Z$ boson, assuming a gluon fusion production mechanism.}
\label{fig:Zgamma}
\end{figure}
Thus, a neutral boson of spin 0, 1, or 2 can be searched for in the $Z\gamma$ channel, allowing for a broad discovery 
program. CMS collaboration performed a search for spin-0 Z$\gamma$ resonances in the leptonic and hadronic decay channels of the $Z$ boson,
on 35.9 fb$^{-1}$ data collected at 13 TeV~\cite{CMS:2017fge}. 
While a search in the leptonic $Z$ boson decay modes has lower SM background, resulting in a higher 
sensitivity for new resonance masses less than about 1 TeV, for higher mass values 
the hadronic $Z$ boson decay channel dominates the sensitivity.
In the hadronic channel, the $Z$ boson candidates are reconstructed by a large-radius light-quark or $b$-quark jet identified using jet substructure techniques.
The hadronic channel separates events into three independent categories, which are created according to number of b-tagged subjects inside the fat-jet, and the
value of N-subjettiness ratio $\tau_{21}$. 
The observed $M_{Z\gamma}$ invariant mass spectra in the signal region ($75<M_{\mathrm{J}}^{\mathrm{pruned}}<105$ GeV) in one of the categories, 
along with the corresponding fits, are shown in Figure~\ref{fig:Zgamma}-left.
No significant deviation is observed and the results in the 
leptonic and hadronic channels are combined and interpreted in terms of upper limits on the production cross section of narrow and broad 
spin-0 resonances with masses between 0.3 and 4.0 TeV, as presented in Figure~\ref{fig:Zgamma}-right.  
A similar search, performed by ATLAS experiment, can be found in Ref.~\cite{ATLAS_poster}.
%

\section{Searches targeting specific particles}
Some searches targeting specific new particles and having more complex final states are discussed in this section.
\subsection{$W' \rightarrow e/\mu + \nu$}
Extensions to the SM may include heavy, charged, gauge boson, which is basically a heavier
version of the SM $W$ boson and is generically referred to as $W'$ boson. The Sequential
Standard Model (SSM) predicts a $W'_{\mathrm{SSM}}$ boson with couplings to fermions that are identical to those of the SM $W$ boson, 
The SSM is useful for comparing the sensitivity of different experiments and is a good benchmark as it allows to reinterpret the results in the context of other new physics models.
A search for $W'_{\mathrm{SSM}}$ boson decaying to an electron or muon and a neutrino using 36.1 fb$^{-1}$ $pp$ collision data at a 
centre-of-mass energy of $\sqrt{s}=13$ TeV is performed by the 
ATLAS collaboration in lepton $+$ missing transverse energy final state~\cite{ATLAS:2017jdr}. 
\begin{figure}[htb]
\centering
\includegraphics[height=2.3in]{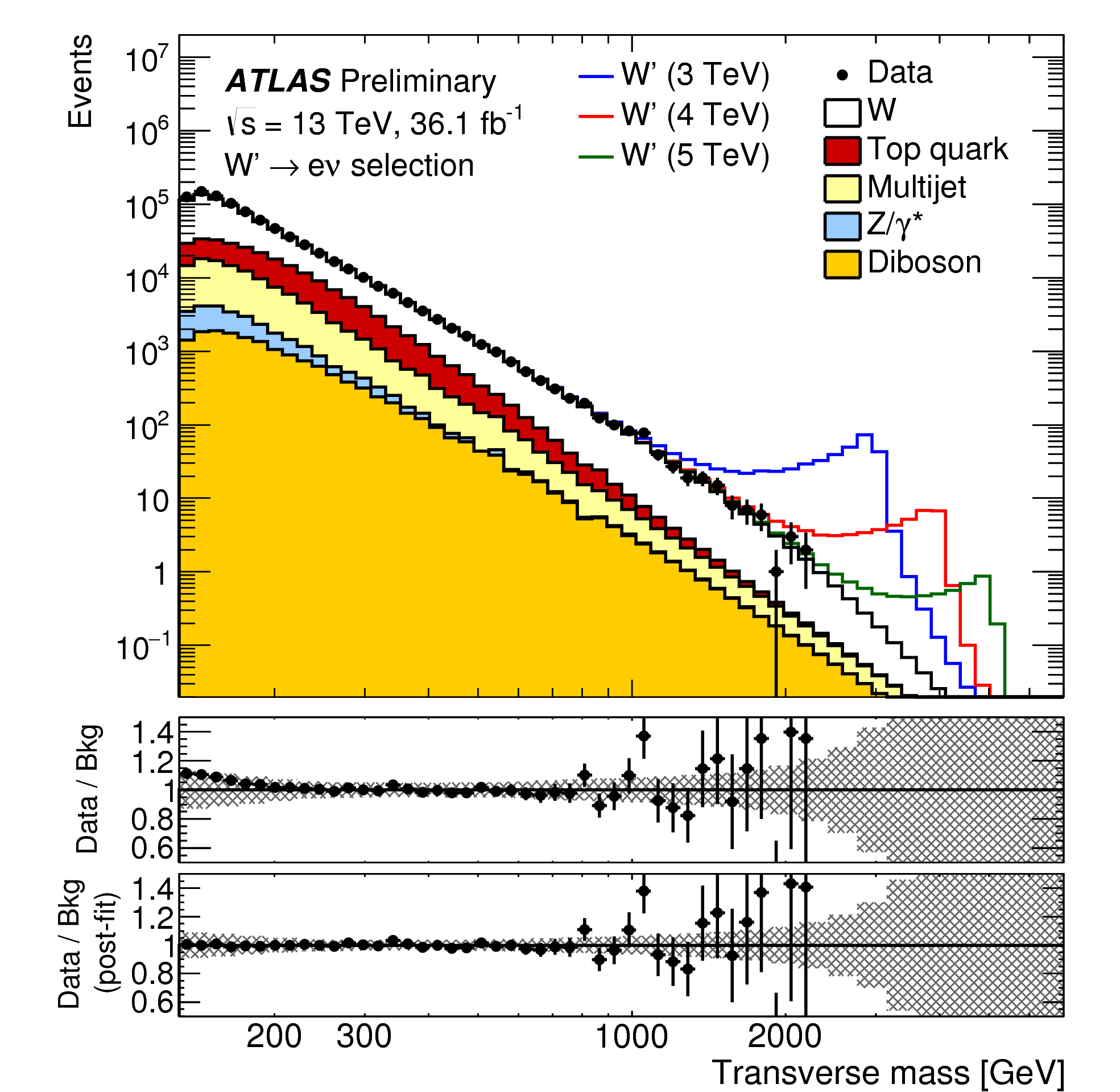}
\hskip 10 pt
\includegraphics[height=2.3in]{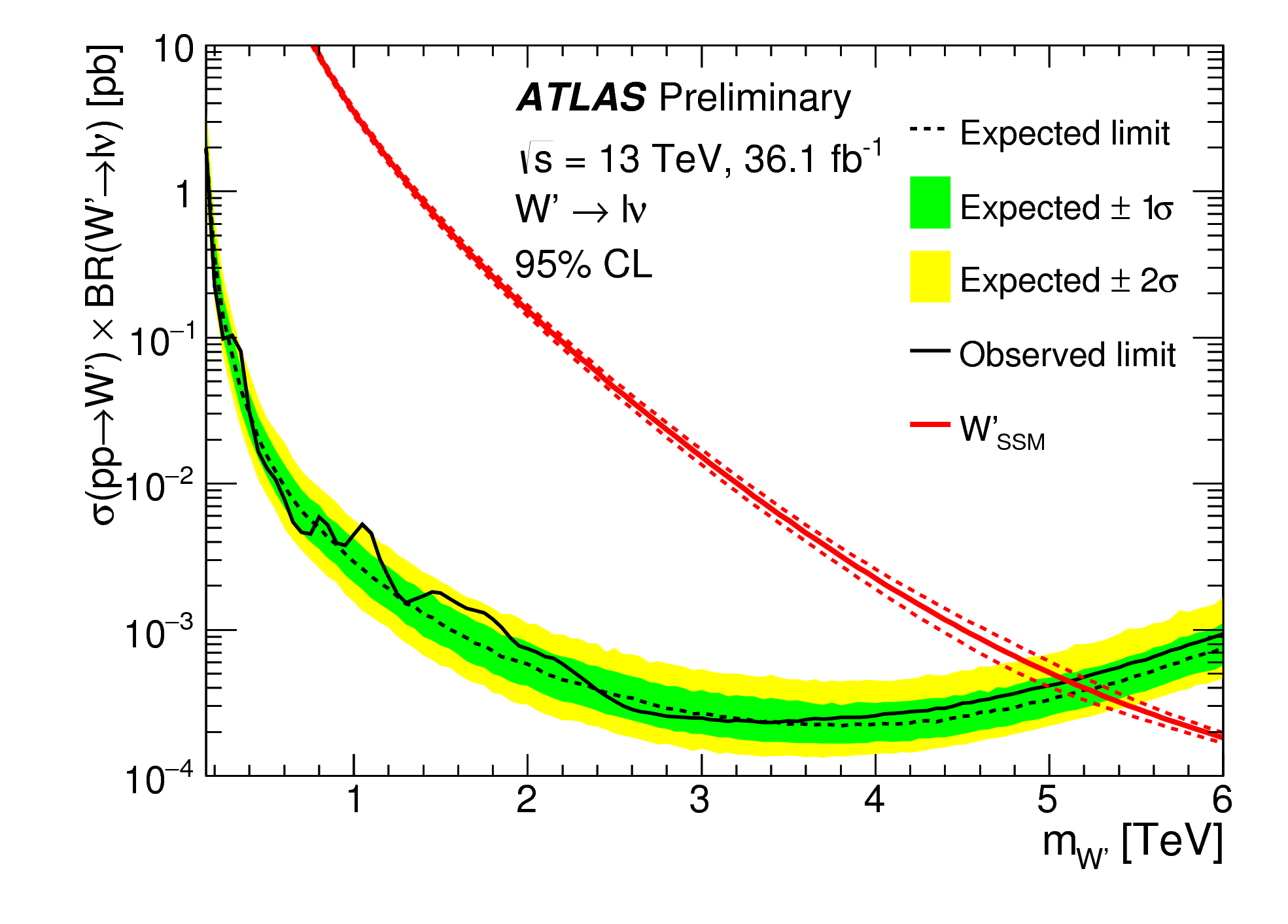}
\caption{Left: Transverse mass distributions~\cite{ATLAS:2017jdr} for events satisfying all selection criteria in the electron $+$ missing transverse energy channel. 
Expected signal distributions for three different $W'$ boson masses are shown.
Right: Observed and expected upper limits~\cite{ATLAS:2017jdr} on cross-section times branching ratio in the combined electron and muon channels. }
\label{fig:wprime}
\end{figure}
The signal discriminant is the transverse mass, which is defined as:
$$
M_{\mathrm{T}}=\sqrt{2 p_{\mathrm{T}} E_{\mathrm{T}}^{\mathrm{miss}} (1 - \cos\phi_{l\nu})}
$$
where $\phi_{l\nu}$ is the angle between the lepton and $E_{\mathrm{T}}^{\mathrm{miss}}$ in the transverse plane.
Examining the $M_{\mathrm{T}}$ spectrum, no significant excess above the expected Standard Model
background is observed. The $M_{\mathrm{T}}$ distribution for electron channel is shown in Figure~\ref{fig:wprime}-left.
$W'_{\mathrm{SSM}}$ with masses below 5.11 TeV is excluded at the 95\% confidence level, as shown in 
Figure~\ref{fig:wprime}-right.
A similar search, performed by the CMS experiment, can be found in Ref.~\cite{Khachatryan:2016jww}.

%
%
\subsection{Heavy right-handed neutrino and $W_R$}
A search for heavy right handed neutrinos $N_l$ ($l= e, \mu$), and right handed $W_R$ bosons, in events with two leptons and two jets, is performed by the CMS experiment using the 
2.6 fb$^{-1}$ of data recorded in 2015 at a center-of-mass energy of 13 TeV~\cite{CMS:2017uoz}. 
\begin{figure}[htb]
\centering
\includegraphics[height=2.3in]{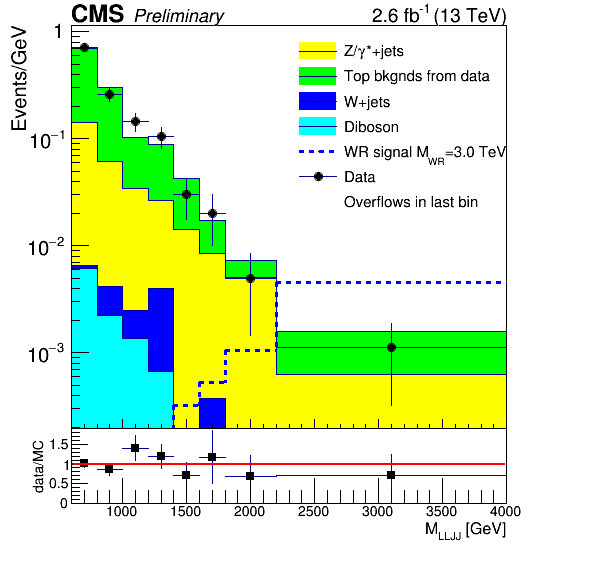}
\hskip 10 pt
\includegraphics[height=2.3in]{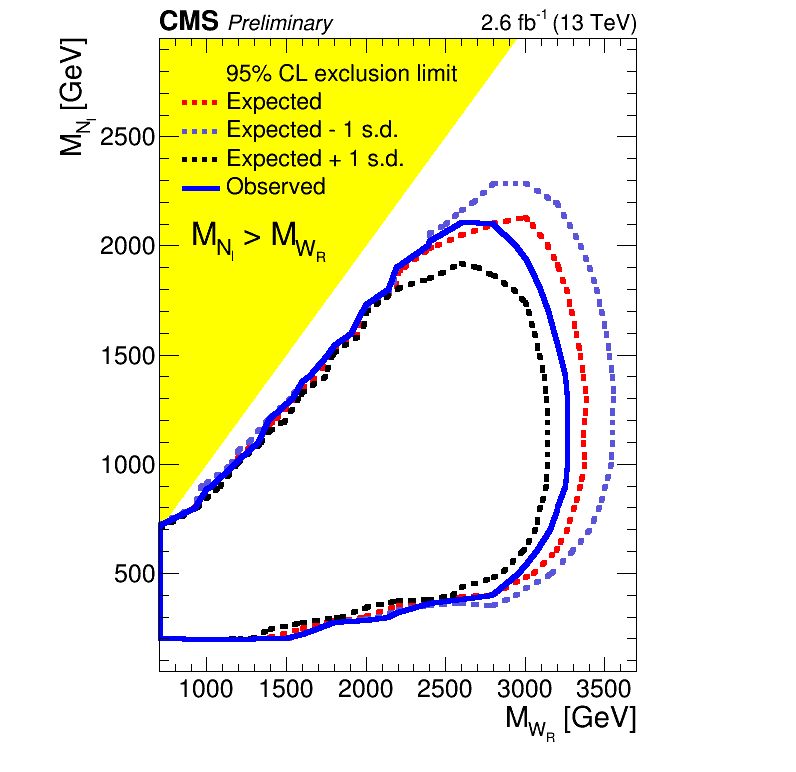}
\caption{
Left: Distribution of $M_{eejj}$~\cite{CMS:2017uoz}. For the $W_R$ signal shown,
$M_{N_l} = \frac{1}{2} M_{W_R}$ . The last bin includes all events above 2200 GeV. 
Right: 95\% confidence level exclusion in the ($M_{W_R}, M_{N_l}$) plane~\cite{CMS:2017uoz} for $W_R \rightarrow eejj$.
}
\label{fig:wr}
\end{figure}
This search assumes the following decay chain of $W_R$ to SM particles:
$$
W_R \rightarrow l_1 N_l \rightarrow l_1 l_2 W_R^* \rightarrow l_1 l_ 2 q \bar{q}'
$$
The only constraint applied to their masses is that $M_{N_l} < M_{W_R}$.
A 2.8$\sigma$ excess was observed by CMS in 8 TeV data in the $eejj$ final state, but not in the $\mu\mu jj$ final state.
In 2015 data, however, no statistically significant excess is found. 
Figure~\ref{fig:wr}-left shows $M_{eejj}$ distribution.
$W_R$ boson production is excluded at the 95\% confidence level 
up to $M_{W_R}=3.5$ $(3.3)$ TeV in the muon (electron) channel.
Figure~\ref{fig:wr}-right shows 95\% confidence level exclusion in the 2D plane of ($M_{W_R}, M_{N_l}$) for $W_R \rightarrow eejj$.
%
%
\subsection{$W' \rightarrow tb$}
A search for heavy gauge bosons decaying to a top and a bottom quarks is performed by CMS collaboration using 35.9 fb$^{-1}$ data at 13 TeV~\cite{CMS:2017rlc}.
\begin{figure}[htb]
\centering
\includegraphics[height=2.3in]{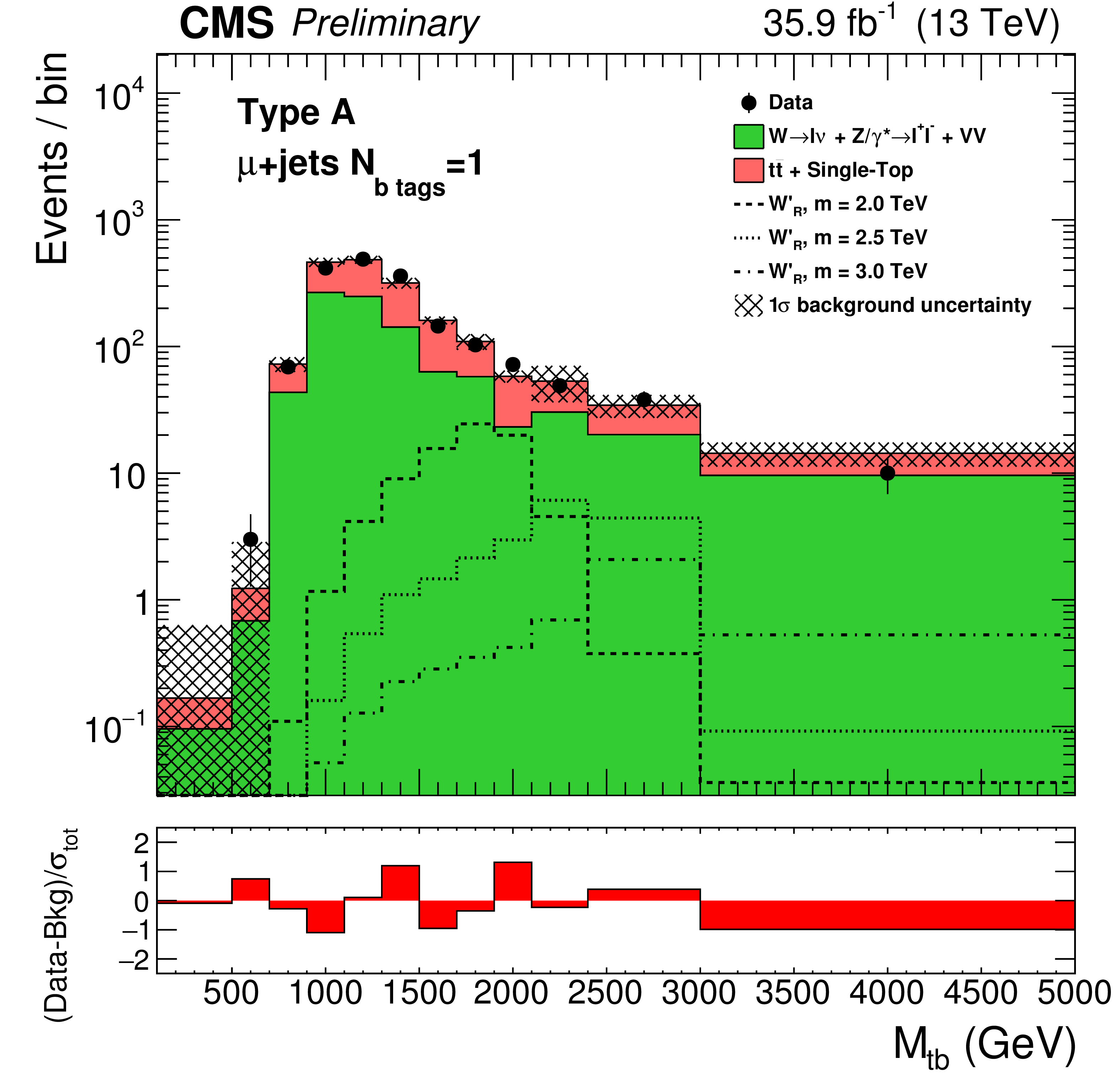}
\hskip 10 pt
\includegraphics[height=2.3in]{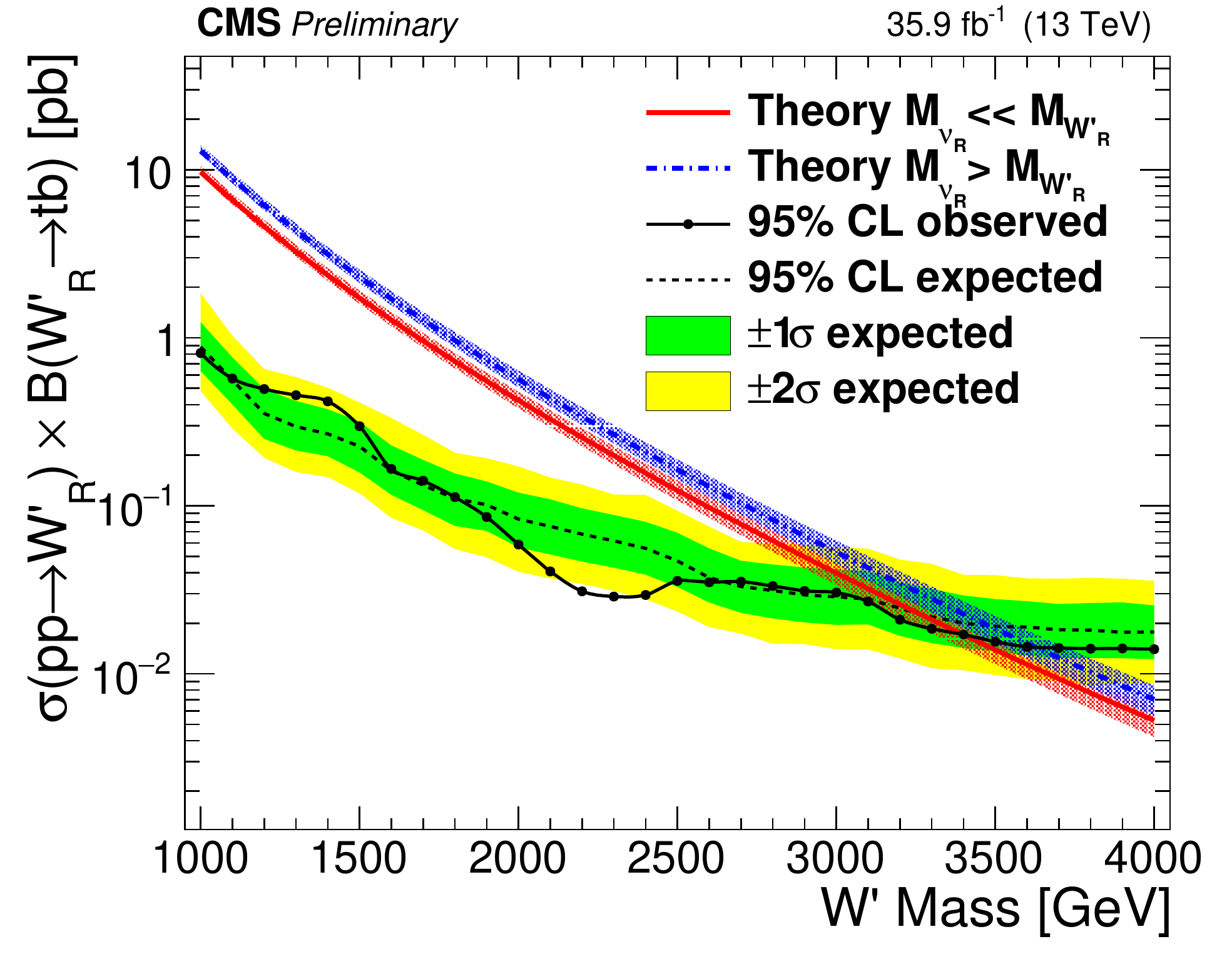}
\caption{
Left: The reconstructed $M_{\mathrm{tb}}$ distribution in the 1 $b$-tag category for the muon channel for Type A events~\cite{CMS:2017rlc}. 
The simulated background and $W'_R$ signal samples are normalized to the cross section and the luminosity of the dataset used. 
Right: 95\% C.L. upper limit~\cite{CMS:2017rlc} on the $W'$ boson production cross section for right-handed $W'$. }
\label{fig:tb}
\end{figure}
Final states that include a single lepton ($e$, $\mu$), multiple jets, and missing transverse energy are analyzed, targeting the decay chain 
$$
W' \rightarrow tb \rightarrow (bl\nu)b.
$$
The search directly probes the $W'$ coupling to third generation quarks, which can be enhanced with respect to lighter quarks in some models.
This search is less impacted by the large continuum multijet background when compared to searches for the decay to light quarks ($W' \rightarrow qq'$).
The analysis separates events into eight independent categories in order to improve the sensitivity of the analysis. 
Categories are created according to lepton type (electron or muon), the number of $b$-tagged jets out of the first two leading $p_{\mathrm{T}}$ jets (=1 or =2), and top $p_{\mathrm{T}}$ and 
dijet $p_{\mathrm{T}}$ (Type A or Type B). 
Events with $p_{\mathrm{T}}^{\mathrm{top}}>650$ GeV and $p_{\mathrm{T}}^{j_1+j_2}>700$ GeV are labelled as Type B events. Events 
which are not labeled Type B are labelled Type A events. 
Comparison between the predicted background and observed data for one of the categories is shown in Figure~\ref{fig:tb}-left.
No evidence for the production of a new W' boson was found.
The production of right-handed $W'$ bosons are excluded at 95\% confidence level for masses below 3.4 TeV if $M_{W_R'} \gg M_{\nu_R}$ and 
upto 3.6 TeV if $M_{W_R'} < M_{\nu_R}$, as shown in Figure~\ref{fig:tb}-right. 



\section{Conclusions}
Latest results on high mass searches performed by the ATLAS and CMS experiments using up to 36.1 fb$^{-1}$ of 13 TeV $pp$ collision data is presented. 
No significant excess have been observed in any of the searches but the reach of the searches are significantly extended with respect to previously published results.
Further searches for new heavy resonances and in more complicated final states are
underway.


\begin{thebibliography}{99}


\bibitem{Chatrchyan:2008aa} 
  CMS Collaboration,
  JINST {\bf 3}, S08004 (2008).


\bibitem{Aad:2008zzm} 
  ATLAS Collaboration,
  JINST {\bf 3}, S08003 (2008).
  
  \bibitem{CMS:2017xrr} 
  CMS Collaboration,
  CMS-PAS-EXO-16-056.

\bibitem{Aaboud:2017yvp} 
  ATLAS Collaboration,
  arXiv:1703.09127 [hep-ex].
  
  \bibitem{ATLAS:2016gvq} 
  ATLAS Collaboration,
  ATLAS-CONF-2016-060.
  
  \bibitem{ATLAS:2016xiv} 
  ATLAS Collaboration,
  ATLAS-CONF-2016-030.
  
    
  \bibitem{CMS:2017dhi} 
  CMS Collaboration,
  CMS-PAS-EXO-17-001.
  
   
    \bibitem{Sirunyan:2017dnz} 
  CMS Collaboration,
  arXiv:1705.10532 [hep-ex].

  
  \bibitem{ATLAS:2016bvn} 
  ATLAS Collaboration,
  ATLAS-CONF-2016-070.

  
  \bibitem{ATLAS:2016sfd} 
  ATLAS Collaboration,
  ATLAS-CONF-2016-084.
  
  \bibitem{CMS:2016pkl} 
  CMS Collaboration,
  CMS-PAS-EXO-16-029.

    
 \bibitem{ATLAS:2017wce} 
  ATLAS Collaboration,
  ATLAS-CONF-2017-027.
  
\bibitem{CMS:2016abv} 
CMS Collaboration,
  CMS-PAS-EXO-16-031.
  
  \bibitem{ATLAS:2016eeo} 
  ATLAS Collaboration,
  ATLAS-CONF-2016-059.
  
  \bibitem{Khachatryan:2016yec} 
  CMS Collaboration,
  Phys.\ Lett.\ B {\bf 767}, 147 (2017)
  [arXiv:1609.02507 [hep-ex]].
  
  \bibitem{CMS:2017fge} 
  CMS Collaboration,
  CMS-PAS-EXO-17-005.
  
  
  \bibitem{ATLAS_poster} 
  ATLAS Collaboration,
  arXiv:1708.00212 [hep-ex].
  
  
 \bibitem{ATLAS:2017jdr} 
  ATLAS Collaboration,
  ATLAS-CONF-2017-016.

\bibitem{Khachatryan:2016jww} 
  CMS Collaboration,
  Phys.\ Lett.\ B {\bf 770}, 278 (2017)
  [arXiv:1612.09274 [hep-ex]].

\bibitem{CMS:2017uoz} 
  CMS Collaboration,
  CMS-PAS-EXO-16-045.
  
  \bibitem{CMS:2017rlc} 
  CMS Collaboration,
  CMS-PAS-B2G-17-010.

\end{thebibliography}
\end{document}